\documentclass[10pt,twocolumn,floatfix,prb,superscriptaddress,]{revtex4-1}
\usepackage{amsmath,amssymb,amsthm,mathrsfs,amsfonts,dsfont,amstext} 
\usepackage{textcomp,pbox}
\usepackage[export]{adjustbox}
\usepackage{bm}
\usepackage{dcolumn,booktabs,url}
\usepackage[scaled]{helvet}
\usepackage{sansmath,gensymb}
\usepackage{tikz,graphicx,transparent,color}
\usepackage{multirow}
\usepackage[separate-uncertainty = true]{siunitx}
\usepackage{comment}
\usepackage{physics}
\usepackage{pdfpages}
\makeatletter
\AtBeginDocument{\let\LS@rot\@undefined}
\makeatother

\usepackage[colorlinks=true]{hyperref}
\usepackage{graphicx}

\hypersetup{
     colorlinks   = true,
     citecolor    = red,
     linkcolor    = blue,
     urlcolor     = red     
}

\graphicspath{{./figs/}}

\newcommand{\yso}{Y$_2$SiO$_5$}
\newcommand{\yb}[0]{Yb$^{3+}$}
\newcommand{\ybisotope}[0]{$^{171}$Yb$^{3+}$}
\newcommand{\ybiso}[0]{$^{171}$Yb$^{3+}$:Y$_2$SiO$_5$}
\newcommand{\y}[0]{Y$^{3+}$}

\newcommand{\figref}[1]{\figurename{~\ref{#1}}}

\newcommand{\mathbbm}[1]{\text{\usefont{U}{bbm}{m}{n}#1}}

\newcommand*{\citen}[1]{%
  \begingroup
    \romannumeral-`\x 
    \setcitestyle{numbers}%
    [\cite{#1}]%
  \endgroup   
}

\begin{document}

\newcommand{\TitleName}{Simultaneous coherence enhancement of optical and microwave transitions in solid-state electronic spins}
\title{\TitleName}

\newcommand{\AffGeneve}{Groupe de Physique Appliqu\'{e}e, Universit\'e de Gen\`{e}ve, CH-1211 Gen\`{e}ve, Switzerland}
\newcommand{\AffParis}{Chimie ParisTech, PSL University, CNRS, Institut de Recherche de Chimie Paris, 75005 Paris, France }
\newcommand{\AffPariss}{Sorbonne Universit\`{e}, Facult\'{e} des Sciences et Ingénierie, UFR 933, Paris, France}

\author{Antonio~Ortu} 
\thanks{These authors contributed equally to this work.}
\affiliation{\AffGeneve{}}
\author{Alexey~Tiranov}
\thanks{These authors contributed equally to this work.}
\affiliation{\AffGeneve{}}
\author{Sacha~Welinski}
\affiliation{\AffParis{}}
\author{Florian~Fr\"owis}
\affiliation{\AffGeneve{}}
\author{Nicolas~Gisin}
\affiliation{\AffGeneve{}}
\author{Alban~Ferrier}
\affiliation{\AffParis{}}
\affiliation{\AffPariss{}}
\author{Philippe~Goldner}
\affiliation{\AffParis{}}
\author{Mikael~Afzelius}\email[Email to: ]{mikael.afzelius@unige.ch}
\affiliation{\AffGeneve{}}

\date{\today}

\begin{abstract}
Solid-state electronic spins are extensively studied in quantum information science, as their large magnetic moments offer fast operations for computing~\cite{Wesenberg2009} and communication~\cite{Togan2010,Gao2013,Bussieres2014}, and high sensitivity for sensing~\cite{Grinolds2014}. However, electronic spins are more sensitive to magnetic noise, but engineering of  their spectroscopic properties, e.g. using clock transitions and isotopic engineering, can yield remarkable spin coherence times, as for electronic spins in GaAs~\cite{Bluhm2011}, donors in silicon~\cite{George2010,Tyryshkin2011,Wolfowicz2013,Lo2015,Morse2017} and vacancy centres in diamond~\cite{Balasubramanian2009,Sukachev2017}. Here we demonstrate simultaneously induced clock transitions for both microwave and optical domains in an isotopically purified \ybiso{} crystal, reaching coherence times of above 100~\textmu s and 1~ms in the optical and microwave domain, respectively. This effect is due to the highly anisotropic hyperfine interaction, which makes each electronic-nuclear state an entangled Bell state. Our results underline the potential of \ybiso{} for quantum
processing applications relying on both optical and spin manipulation, such as optical quantum memories~\cite{Bussieres2014,Saglamyurek2015a}, microwave-to-optical quantum transducers~\cite{Probst2013,Fernandez-Gonzalvo2015}, and single spin detection~\cite{Bienfait2016}, while they should also be observable in a range of different materials with anisotropic hyperfine interaction.
\end{abstract}

\maketitle 

A very effective method for increasing spin coherence times consists in engineering a particular spin transition so that its frequency $\nu$ is insensitive to magnetic fluctuations, i.e. $\partial \nu / \partial \vec{B} = 0$ up to first order. Such transitions are known as clock transitions in atomic physics, and were pioneered for nuclear spin systems in solid-state materials \cite{Fraval2004} where they are called ZEFOZ (ZEro First Order Zeeman) transitions. The ZEFOZ technique is based on the application of a magnetic bias field in a particular "magic" direction, where $\partial \nu / \partial \vec{B} = 0$ simultaneously for all three spatial coordinates. The existence of such a direction depends entirely on the spin Hamiltonian of the system. Recently ZEFOZ transitions were also observed for electronic spins in silicon \cite{Wolfowicz2013,Morse2017}, showing its general appeal for spin coherence enhancement.

In rare-earth (RE) ion doped crystals, engineered ZEFOZ transitions have been very effective in extending nuclear spin coherence times, by orders of magnitude, for REs without electronic spin ($S = 0$) \cite{Fraval2004,Lovric2011,Zhong2015}. 
However, ZEFOZ transitions have not yet been achieved in RE ions having both electronic and nuclear spin degrees of freedom, so-called Kramers RE ions (containing odd number of electrons) \cite{Bertaina2007,Wolfowicz2015,Rancic2017}. 
These often have shorter coherence lifetimes due to the large electronic spin coupling to the spin bath of the crystal, which is composed of nuclear spins and the RE spins themselves, unless large magnetic fields are applied \cite{Rancic2017}. In terms of optical coherence times, RE ions are unique solid state centres as optical coherences can reach milliseconds \cite{Thiel2011b}. In short this is due to the shielding of the 4$f$ electrons due the 5$s$, 5$p$  orbitals, which reduces the coupling to phonons in RE materials. But also in the optical domain one needs to apply large magnetic fields to reach long optical coherence times for Kramers ions \cite{Sun2002}. Yet, electronic spins are highly promising for broadband optical quantum memories \cite{Bussieres2014,Saglamyurek2015a}, microwave-to-optical quantum transduction and coupling to superconducting qubits \cite{Probst2013,Fernandez-Gonzalvo2015}. There is thus a general interest in simultaneously enhancing the spin and optical coherence times of Kramers ions, in order to profit from their distinct advantages. In addition, to interface with superconducting resonators and qubits, ZEFOZ transitions under low or zero magnetic fields are of particular interest.

\begin{figure}
	\includegraphics[width=\linewidth]{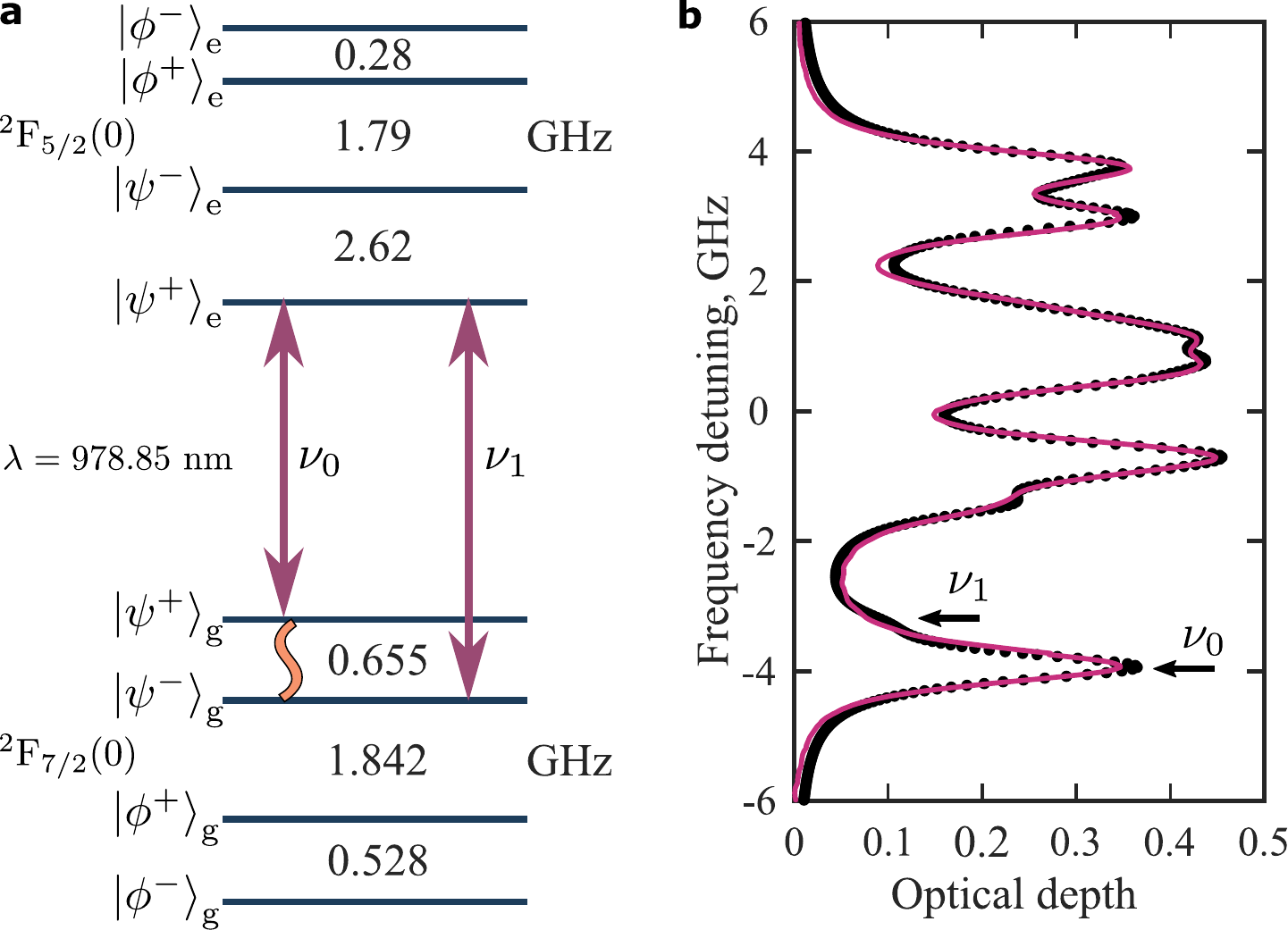}
	\caption{(color online) \textbf{Energy levels and absorption profile}. \textbf{a} Energy level diagram for the optical transitions of \ybiso{} under study and  frequencies of the hyperfine transitions for the ground and excited states. Optical transitions identified with $\nu_0$ and $\nu_1$ are utilized to study the spin and optical coherences. \textbf{b} Absorption spectrum at zero magnetic field. Data is shown in black, while the pink line 
represents the result from the fitting based on the zero field energy structure~\cite{SpectroYb}. The black arrows indicate the optical transitions at $\nu_0$ and $\nu_1$ used in this work. }
	\label{fig:zero_field_ab}
\end{figure}

In this study we use $^{171}$Yb$^{3+}$, a Kramers ion that recently has received interest for quantum information processing owing to its simple hyperfine structure. Basic optical \cite{Bottger2016,Welinski2016,SpectroYb} and spin \cite{Welinski2016,SpectroYb} properties have been studied, and recent electron spin resonance measurements have shown promising spin coherence times at high magnetic fields (73~\textmu s at around 1~T~\cite{Lim2018a}). The $^{171}$Yb$^{3+}$ is indeed unique among RE Kramers ions, as it has the lowest possible, non-zero, electronic $S = \frac{1}{2}$ and nuclear $I = \frac{1}{2}$ spin, resulting in the simplest possible hyperfine manifold of four states, as shown in \figref{fig:zero_field_ab}a. This fact greatly simplifies both the optical  and spin spectra and makes optical manipulation of the hyperfine levels possible, an important feature of this work.

\begin{figure}
	\includegraphics[width=\linewidth]{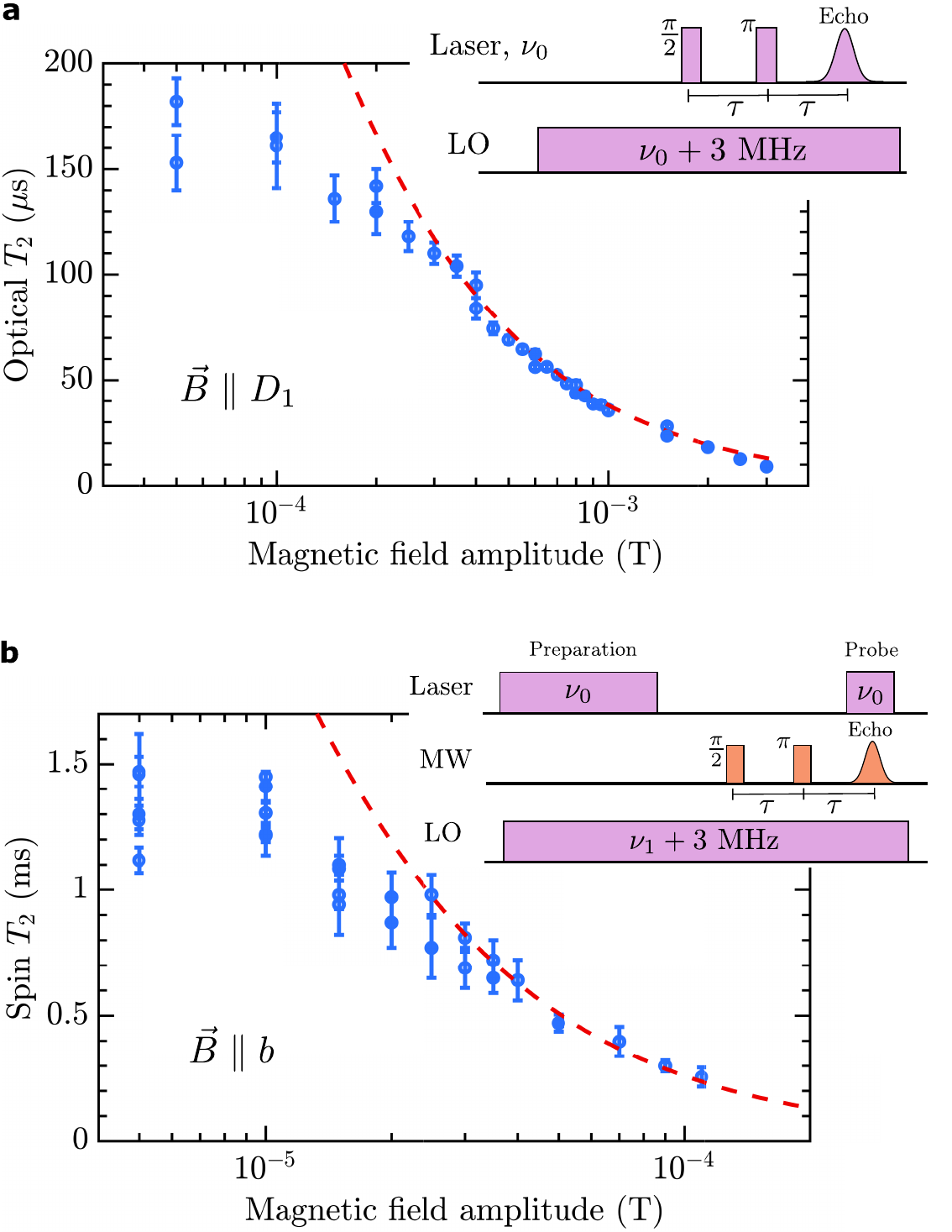}
	\caption{(color online)  \textbf{Coherence enhancement under zero magnetic field.} \textbf{a} Experimental photon echo sequence and coherence lifetime $T_2^\text{(o)}$ measurements as a function of the external magnetic field. \textbf{b} Experimental pulse sequence used to study spin coherence on 655~MHz hyperfine transition. The preparation step produces a population difference between states $\ket{\psi^{+}}_g$ and $\ket{\psi^{-}}_g$ before applying series of microwave (MW) pulses to create the spin echo sequence. The echo signal is measured using Raman heterodyne scattering by detecting the beating between the coherently scattered emission at $\nu_1$ and the local oscillator (LO).  The results for the spin coherence lifetime $T_2^\text{(s)}$ as a function of the external magnetic field are shown on the right. Dashed lines represent theoretical curves based on the decoherence model (see main text for details). The error bars correspond to 95\% confidence intervals on the fitted decay of the echoes (see methods). Note that for some magnetic fields, particularly at extremely low fields, multiple data points are shown in panels \textbf{a} and \textbf{b}. These were measured during different days and generally show a good repeatability of the experiment. The relatively larger data spread for the lowest fields we attribute to small day-to-day variations of the bias field of the lab.}
	\label{fig:zero_field_cd}
\end{figure}

\begin{figure*}
	\includegraphics[width=\linewidth]{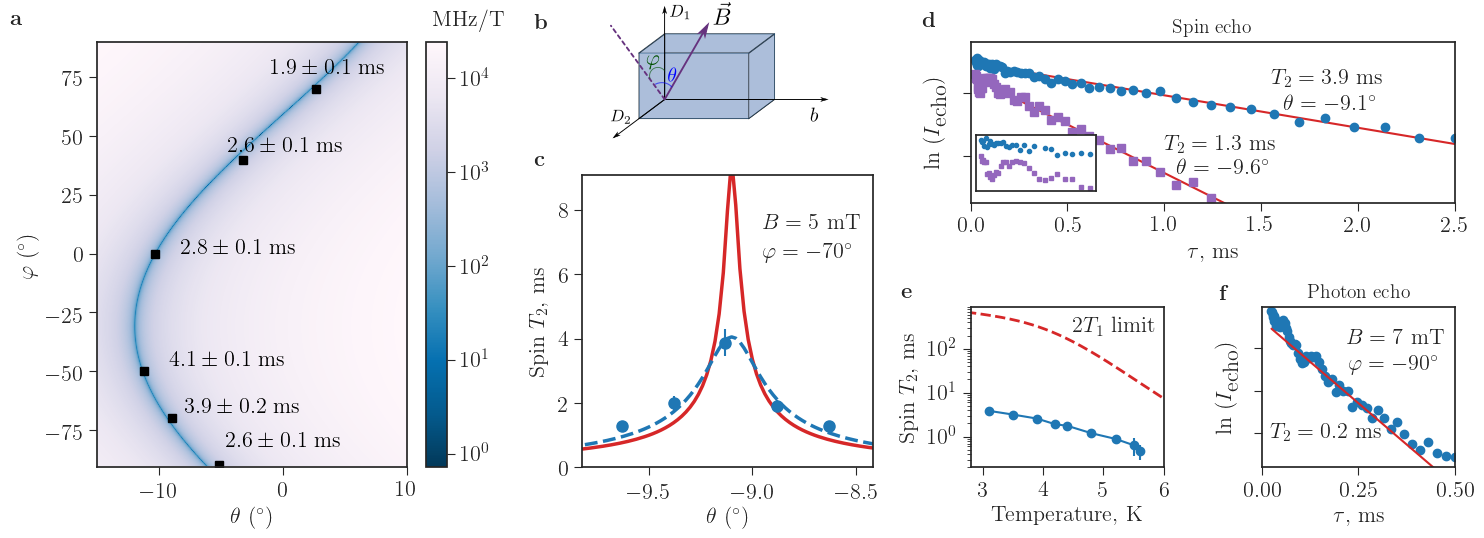}
	\caption{(color online) \textbf{Low magnetic field assisted spin coherence enhancement}. \textbf{a} Gradient of the spin transition for ground state of \ybiso{} for external magnetic field of 5~mT applied in different directions. $(\theta,\varphi)$ are spherical angular coordinates in the ($\mathbf{D_1}$, $\mathbf{D_2}$, $\mathbf{b}$) frame. The dark region  represents the directions where the gradient $\bf{S_1}$ is minimal. Some examples of the measured values of $T_2^{(\text{s})}$ are shown at their respective coordinates. \textbf{b} Scheme showing the direction of the magnetic field with respect to the axes $\mathbf{D}_1$, $\mathbf{D}_2$ and $\mathbf{b}$ of the \yso{} crystal. \textbf{c}~Spin coherence lifetime measured for $\varphi = -70 ^{\circ}$ as a function of $\theta$ near the region with minimal decoherence (points). While the theoretical calculation (solid line) predicts bigger coherence lifetimes, the results can be explained by the magnetic field inhomogeneity inside the crystal (see discussion in the main text). 
\textbf{d}~Two-pulse spin echo intensity plotted against the delay between the microwave excitation pulses. Circles show the decay corresponding to the optimal point in \textbf{c}. Squares show the decay with the magnetic field tilted by 0.5${^\circ}$ from the optimum. The inset shows a zoom of the first 300 $\mu$s, where the decay displays oscillations due to coupling to nuclear spins in the host crystal. \textbf{e} The optimized spin coherence lifetime $T_2^{(\text{s})}$ measured as a function of the crystal temperature. The dashed line shows the limit $2 T_1^{(\text{s})}$ due to the spin-lattice relaxation $T_1^{(\text{s})}$ lifetime measured in Ref. \cite{Lim2018a}. \textbf{f} An example of a photon echo decay measurement for a magnetic field of 7 mT. The error bars on $T_2$ in all figures (most of the time smaller than the symbol representing the data point) correspond to 95\% confidence intervals on the fitted decay of the echoes (see methods).}
	\label{fig:magic_angle}
\end{figure*}

In this system, the electronic spin $\mathbf{S}$ of the ytterbium ion is coupled with its nuclear spin $\mathbf{I}$ through the hyperfine interaction tensor~$\mathbf{A}$, and the effective spin Hamiltonian involving the interaction with an external magnetic field $\mathbf{B}$ can be written as
\begin{equation}
\label{eq:Heff}
\mathcal{H} = \mathbf{S} \cdot \mathbf{A} \cdot \mathbf{I} + \mu_\text{B} \mathbf{B} \cdot \mathbf{g} \cdot \mathbf{S} - \mu_\text{n}  \mathbf{B}\cdot \mathbf{g}_\text{n} \cdot \mathbf{I}.
\end{equation}
Here, $\mathbf{g}$ and $\mathbf{g}_\text{n}$ are coupling tensors related  to the electronic and nuclear Zeeman interactions of \ybisotope{}, respectively,  and $\mu_B$ and $\mu_n$ are the electronic and nuclear magnetons. In the \yso{} crystal the $^{171}$Yb$^{3+}$ ions replace Y$^{3+}$ ions in sites of low ($C_1$) point symmetry, which makes the hyperfine $\mathbf{A}$ and Zeeman $\mathbf{g}$ tensors highly anisotropic for both the ground and excited states~\cite{Welinski2016}. In general, for a highly anisotropic $\mathbf{A}$ tensor (with non-zero eigenvalues $A_x \neq A_y \neq A_z$)), the hyperfine coupling completely removes the degeneracy at zero magnetic field. It gives rise to the four eigenstates  $\ket{\psi^{\pm}} = \frac{1}{\sqrt{2}}(\ket{\uparrow \Downarrow} \pm \ket{\downarrow \Uparrow})$ and  $\ket{\phi^{\pm}} =  \frac{1}{\sqrt{2}}(\ket{\uparrow \Uparrow} \pm \ket{\downarrow \Downarrow})$, where we decompose electronic $\ket{\uparrow} \equiv \ket{S_z = \frac{1}{2}}$, $\ket{\downarrow} \equiv \ket{S_z = -\frac{1}{2}}$ and nuclear $\ket{\Uparrow} \equiv \ket{I_z = \frac{1}{2}}$, $\ket{\Downarrow} \equiv \ket{I_z = -\frac{1}{2}}$ spin components. Their energies are expressed as  $E_{\psi^{\pm}} =  \frac{1}{4}\left[ A_z \pm (A_x - A_y) \right]$ and $E_{\phi^{\pm}} =  \frac{1}{4} \left[ - A_z \pm (A_x + A_y) \right]$.

By tracing out the electronic (nuclear) spin one can easily see that the resulting density matrix corresponds to the complete mixture of two orthogonal  components $\rho_e = \frac{1}{2} \dyad{\uparrow}{\uparrow} + \frac{1}{2}\dyad{\downarrow}{\downarrow}  =  \mathbbm{1}/2$ ($\rho_n = \frac{1}{2}\dyad{\Uparrow}{\Uparrow}  + \frac{1}{2}\dyad{\Downarrow}{\Downarrow}=  \mathbbm{1}/2$), which is true for any eigenstate. This inherently gives a zero total spin vector for the electronic $\expval{\mathbf{S}} = 0$ (nuclear $ \expval{\mathbf{I}} = 0$) spin polarization. Hence any first-order perturbation of the energy levels by the Zeeman terms of Eq.~\eqref{eq:Heff} is strictly zero for both the electronic and nuclear Zeeman components. More explicitly, for any state $\ket{\xi} = \ket{\psi^\pm}, \ket{\phi^\pm}$ the following is true:
$\bra{\xi} \mathbf{B}_\text{p} \cdot \mathbf{g} \cdot \mathbf{S} \ket{\xi}  = \bra{\xi} \mathbf{B}_\text{p} \cdot \mathbf{g_n} \cdot \mathbf{I} \ket{\xi} = 0$ for any perturbing magnetic field $\mathbf{B}_\text{p}$. These results apply to any electronic state, hence the zero-field ZEFOZ condition occurs for any spin transition in the ground and excited states, and for any optical transition connecting the two electronic states. A similar effect has been observed in the hyperfine ground state in NV centres in diamond~\cite{Dolde2011,Jamonneau2016}, but only for a reduced number of states due to the higher symmetry of the interaction tensor and particularly not for any optical transition.

To probe the zero-field spin and optical coherences in \ybiso{} we use the transitions $\ket{\psi^+}_g \leftrightarrow \ket{\psi^-}_g$ and $\ket{\psi^+}_g\leftrightarrow\ket{\psi^+}_e$, the latter being clearly resolved in the absorption spectrum in \figref{fig:zero_field_ab}b. By means of optical pumping on the optical transition, the population difference between two spin states is created. A Hahn sequence on the spin transition follows, inducing an echo pulse in the microwave range, which is optically detected using Raman heterodyne scattering~\cite{Fraval2004} (see Methods for details). The coherence lifetime $T_2^{(\text{s})}$ is thus measured as a function of the external magnetic field applied in various directions.
The optical coherence lifetime $T_2^{(\text{o})}$ is instead measured through a standard photon echo technique using the same optical absorption line (\figref{fig:zero_field_ab}b).

Figure~\ref{fig:zero_field_cd}b shows measurement results for the $T_2^\text{(s)}$ lifetime as the magnetic field is varied close to the expected zero-field ZEFOZ point along the crystal $\mathbf{b}$ axis.
The coherence increases as the magnetic field is lowered, exceeding $\SI{1}{\milli\second}$ lifetime for few tens of $\mu$T. A similar behaviour is observed for the optical coherence $T_2^\text{(o)}$, which approaches $\SI{200}{\micro\second}$ close to zero field (\figref{fig:zero_field_cd}a). These results clearly demonstrate the effectiveness of the zero-field ZEFOZ point for decoupling the $^{171}$Yb spins from the perturbing spin bath. Remarkably these spin and optical coherence times are similar to those found in non-Kramers RE ions (with $S = 0$) at zero field \cite{Thiel2011b}.

The decoherence due to a given magnetic field noise $\mathbf{\Delta B}$ can be modelled as \cite{Zhong2015} $(\pi T_2)^{-1} = \mathbf{S}_1 \cdot \mathbf{\Delta B} + \mathbf{\Delta B} \cdot \mathbf{S}_2 \cdot \mathbf{\Delta B}$, where $\mathbf{S}_1$ and $\mathbf{S}_2$ are the linear (gradient) and quadratic (curvature) Zeeman contributions to a transition energy. In a ZEFOZ point the linear term ideally vanishes and the limitation comes from the quadratic term. Here we use the known hyperfine $\mathbf{A}$ \cite{SpectroYb} and Zeeman~$\mathbf{g}$~\cite{Welinski2016} tensors to calculate $\mathbf{S}_1$ and $\mathbf{S}_2$, and fit the noise term  $\mathbf{\Delta B}$ for fields away from the ZEFOZ point, using both spin and optical data (see \figref{fig:zero_field_cd}a and b). This results in $\abs{\mathbf{\Delta B}} \sim \SI{3}{\micro\tesla}$, consistent with previous measurements in Y$_2$SiO$_5$ and is due to spin flip-flops of yttrium ions in this crystal~\cite{Lovric2011, Zhong2015}. At zero field, however, the model predicts up to $\sim \SI{10}{\milli\second}$ spin coherence time. The lower experimental value  (maximum 4~ms was measured) could possibly be attributed to the presence of a bias magnetic field, which a separate measurement estimated to be $\lesssim \SI{15}{\micro\tesla}$.

We also explore a second ZEFOZ-like regime, using the low-field assisted coherence enhancement technique, as shown on \figref{fig:magic_angle}. It is based on the fact that the direction of an extremely small magnetic bias field can strongly reduce the linear term $\mathbf{S}_1$. This is due to the high anisotropy of the $\mathbf{g}$ tensor for the ground state which has the strongest component $g_z = 6.06$ close to the $\mathbf{b}$ axis of the crystal~\cite{Welinski2016}, while its perpendicular components are up to fifty times lower and vary between $g_x = 0.13$ and $g_y = 1.50$. In addition, the $\mathbf{A}$ tensor is almost parallel to the $\mathbf{g}$ tensor and the two have a similar anisotropy \cite{SpectroYb}. For a bias small field applied in the $x$ direction, the gradient is $|\mathbf{S}_1| \approx  2 B \mu_\text{B}^2   g_x^2 A_y / A_z^2$. For \ybiso{} this results in $|\mathbf{S}_1| \sim 0.001\,\mu_\text{B}$ for bias fields of a few mT along the $x$ axis (see Methods for details). In fact, any field in the $x$-$y$ plane results in a strongly reduced gradient, of the order of a nuclear spin sensitivity. 

Figure~\ref{fig:magic_angle}a shows the numerically calculated $\mathbf{S}_1$ gradient for a range of orientations of the bias field $\vqty{\mathbf{B}}=\SI{5}{\milli\tesla}$. We here work in the ($\mathbf{D_1}$, $\mathbf{D_2}$, $\mathbf{b}$) crystal frame commonly used in optics \cite{Welinski2016}, in which the $x$-$y$ plane is slightly tilted. The calculated gradient varies over four orders of magnitude in the narrow range shown in \figref{fig:magic_angle}a, with minimum gradients lower than 10~MHz/Tesla close to the $x$-$y$ plane. Such gradients are typical of nuclear spins, while electronic spins have gradients of 10~GHz/T or more. The calculations also indicate that a high precision of the bias field alignment is required to reach the minimum region of gradients, as explained below.

Although magnetic sensitivities of the energy levels are on the order of nuclear moment, the dipole moments of the addressed transitions remain as high as for electronic ones. Explicitly, $\bra{\xi^\pm } \mathbf{B}_\text{ac} \cdot \mathbf{g} \cdot \mathbf{S} \ket{\xi^\mp } \propto g \,\mu_\text{B}$ with $\ket{\xi} = \ket{\psi}, \ket{\phi}$ and $g$ being a constant arising from the elements of the electronic $\mathbf{g}$ tensor in the $z$ direction. By aligning $ \mathbf{B}_\text{ac}$ along the strongest component of the $\mathbf{g}$ tensor we achieve Rabi frequencies of around 1 MHz for a weak excitation field of around 30~$\mu$T (see Methods for details).

An example of a spin coherence lifetime measurement taken with a small scan of $\theta$ at $\varphi=\ang{-70}$ is shown in~\figref{fig:magic_angle}c, together with the theoretical model using the calculated gradient and a magnetic noise of $\abs{\mathbf{\Delta B}} \sim \SI{3}{\micro\tesla}$. The peak coherence time is 4~ms, and less than a degree misalignment significantly reduces the coherence time. This confirms that even extremely small bias fields can have a strong effect on the coherence properties in electronic spin systems characterized by a strongly anisotropic $\mathbf{g}$ tensor. Measurements made at different $\varphi$ angles displayed the same behaviour, and the maximum coherence lifetimes are shown at their corresponding angular coordinates in~\figref{fig:magic_angle}a. In the entire plane of minimum gradients we achieve coherence times above 1~ms for a bias field of 5~mT.  

The main limitation in the achieved spin coherence times using the bias field method probably stems from the inhomogeneity of the applied field. As seen in \figref{fig:magic_angle}c, the field has to be precise to within 1/10 of a degree to reach the theoretical maximum. Our numerical calculations show that field inhomogeneities of $\approx 0.5 \%$ in one cm$^3$ volume are enough to explain the experimental data (\figref{fig:magic_angle}c). Ohter possible limitations are discussed below.

Optical coherence times of up to 200 $\mu$s were also measured with magnetic fields applied in the same directions where spin coherence times are maximum. An example of a photon echo decay measurement is shown in \figref{fig:magic_angle}f. The angular dependence was much less pronounced as with respect to the spin coherence measurement, a few degrees misalignment did not significantly reduce the coherence time. Probably, this is due to the dependence of the optical gradient on the $\mathbf{g}$ tensors in both the ground and excited states, which are not completely aligned~\cite{Welinski2016}. Generally this helps one to simultaneously reach long optical and spin coherence times.

Another possible limitation to the achieved coherence times, both in the zero and low field regimes, comes from the superhyperfine interaction taking place between the electronic Yb$^{3+}$ spin and nuclear spins of the environment. In the \yso{} host nearest-neighbour nuclear spins of $^{89}$Y (100\% with $I=\frac{1}{2}$) and $^{29}$Si (5\% with $I=\frac{1}{2}$) interacts strongly with Yb$^{3+}$, and we expect this interaction to modify the transition gradients. Further analysis is required in order to estimate these contributions to the observed spin coherence times. In addition, these interactions cause splitting of the microwave transitions when a magnetic bias field is applied, which causes characteristic oscillations on the spin echo decays curves. Such oscillations were seen at short time scales in the measurements with a small bias field, as shown in~\figref{fig:magic_angle}d. The period of the oscillations is consistent with coupling to $^{89}$Y ions. Surprisingly, these oscillations were systematically quenched in amplitude as the field was tuned to the optimal angle. Although this effect is not yet fully understood, this is an useful feature of these optimal points for applications such as quantum memories.

Finally, we studied the spin coherence lifetime as a function of temperature, with a magnetic field applied in the optimum point for $\varphi=\ang{-70}$. As shown in~\figref{fig:magic_angle}e, the coherence time remains above 100 $\mu$s for temperatures up to 5.6~K. Also, it remains well below the limit given by the spin-lattice relaxation lifetime as measured in Ref. \citen{Lim2018a}. However, cross-relaxation between \ybisotope{} ions are likely to reduce the $T_1$ lifetime of the hyperfine states \cite{Cruzeiro2017}, in addition to causing spectral diffusion \cite{Thiel2011b}. Further studies should measure and quantify these contributions to the measured spin coherence lifetimes, for instance by varying the \ybisotope{} doping concentration and/or applying a magnetic field gradient as shown with naturally doped silicon samples \cite{Tyryshkin2011}.

The spin coherence times we measured are similar to those found in naturally doped silicon, where the nuclear spin of $^{29}$Si causes magnetic noise. In silicon, coherence times of 10 seconds have been reached by isotope purification (pure $^{28}$Si samples) \cite{Tyryshkin2011}. In contrary, yttrium has the unique isotope $^{89}$Y which has a nuclear spin, making such an approach impossible for \yso{}. A possible solution is to use an isotopically enriched versions of tungsten oxide crystals (for example CaWO$_4$), whose naturally doped version was studied with various rare-earth dopants in the past~\cite{Bertaina2007, Rakhmatullin2009}.

The described coherence enhancement techniques can generally be applied to any system having anisotropic Zeeman and hyperfine interaction. Particularly, they can be extended to higher odd nuclear spins ($I = \frac{3}{2}$, $\frac{5}{2}$, $\frac{7}{2}$) coupled with half electronic spin ($S=\frac{1}{2}$) (see Methods for details). Our approach in this case is directly applicable to other isotopes of Kramers ions with non-zero nuclear spin, in particular to $^{167}$Er$^{3+}$ which is of high interest for quantum communication owing to its optical transition in the telecommunications band \cite{Rancic2017}. Recent experiments \cite{Rakonjac2018} in $^{167}$Er$^{3+}$ have reached promising hyperfine coherence times (up to 300~\textmu s) by also using the mixing of electronic and nuclear spins at zero magnetic field, but without fully exploiting a ZEFOZ point. Hence, with ZEFOZ points or the low-field assisted enhancement techniques used in our work, the spin coherence time could potentially be increased further. Other systems of interest include single molecular electron magnets incorporating rare earth ions~\cite{Zadrozny2015,Shiddiq2016}.

In conclusion, we have shown that simultaneous enhancement of both optical and spin coherence times is possible, by using induced clock transitions at zero field or by minimizing the transition gradients using extremely low magnetic bias fields. These techniques are applicable to any electronic spin system having anisotropic Zeeman and hyperfine interactions. The \ybiso{} material studied here features a set of unique properties, such as a simple hyperfine manifold, optically resolved optical-hyperfine transitions and long coherence times, making it a great resource for quantum information applications. This system will be highly interesting for applications in broadband optical quantum memories \cite{Bussieres2014,Saglamyurek2015a} and coupling to superconducting resonators/qubits in the microwave regime \cite{Probst2013,Bienfait2016}.

\section*{Methods}
\subsection*{Crystal}
Our sample is a \ybiso{} crystal with $10$ ppm doping concentration and with an \yb{} isotope purity of $95 \% $, grown via Czochralski method and cut along the $\mathbf{D_1}$, $\mathbf{D_2}$ and $\mathbf{b}$ extinction axes. The sides parallel to these axes are long respectively $5.7$, $5.6$ and $\SI{9.5}{\milli\metre}$, and the faces parallel to the $\mathbf{D_1}-\mathbf{D_2}$ plane were polished so to reduce the scattering of the input light. Another crystal grown from the same bulk but with dimensions $10.0$, $4.6$ and $\SI{4.5}{\milli\metre}$ respectively was polished along the $\mathbf{D_2}-\mathbf{b}$ faces and used only for coherence lifetime measurements on site I (see "Opically detected spin echo" section).
\yso{} has a monoclinic structure of the $C_{2h}^6$ space group. \yb{} can replace \y{} ions in two different crystallographic sites of $C_1$ point symmetry (referred to as \emph{site I} and \emph{II}), which in turn include two magnetically inequivalent sub-sites each. All results presented in the main text were obtained from site II.
The hyperfine transitions of the two magnetic sites coincided when no static magnetic field was applied, but they could be addressed separately when we introduced few $\SI{}{\milli\tesla}$ in directions not orthogonal nor parallel to $\mathbf{b}$.
The optical transitions had a Full Width Half Maximum of $\sim \SI{500}{\mega\hertz}$, while for the hyperfine transitions the FWHM was about $\SI{1}{\mega\hertz}$.

\subsection*{Experimental setup}
The crystal was placed in a cryostat at about $\SI{3.5}{\kelvin}$. A small coil of copper wire, connected to a generator and a $\SI{2}{\watt}$ amplifier, surrounded the crystal along the $\mathbf{b}$ axis and allowed us to address the microwave transitions by inducing ac magnetic fields. 
The inductance of the coil was estimated to be 0.2~\textmu H, which gives $\approx 0.5$~\% of input microwave power to be transmitted to the crystal for 655~MHz excitation frequency. The induced ac magnetic field amplitude in this case could go up to $\approx 30$~\textmu T. 

A larger superconducting coil was placed along either $\mathbf{D_1}$ or $\mathbf{D_2}$ axes, and was used to create a component of the external static magnetic field up to $\SI{2}{\tesla}$. Placed out of the cryostat along the remaining crystal axes, two pairs of copper coils in a Helmholtz configuration completed the static field generation setup, with a maximum of few $\SI{}{\milli\tesla}$ each. This solution proved to be enough for generating the required field along the directions of maximal suppression of decoherence presented in the main text. A $\SI{980}{\nano\metre}$ external cavity diode laser generated the optical beams, which were manipulated in amplitude and frequency by acousto-optical modulators so to precisely define the pulse sequences we needed. The laser was split in two main paths, namely the pump/probe sent on the crystal and a local oscillator used in the heterodyne detection scheme. A non-linear crystal-based phase modulator controlled the frequency of the local oscillator.

\subsection*{Optically detected spin echo}
We measure the spin coherence lifetime $T_2^\text{(s)}$ through optical detection of a spin echo in a Hahn sequence using the Raman heterodyne scattering (RHS) technique \cite{Fraval2004,Lovric2011,Zhong2015}. All spin echo measurements reported in this paper were carried out on the $\ket{\psi^-}_{g}\rightarrow\ket{\psi^+}_{g}$ transition ($\SI{655}{\mega\hertz}$) of optical site II. We also obtained similar results ($T_2^\text{(s)}=\SI{2.4}{\milli\second}$) for the transition $\ket{\phi^-}_{g}\rightarrow\ket{\phi^+}_{g}$ ($\SI{528}{\mega\hertz}$) for a small scan of magnetic field along $\mathbf{D_2}$ in the range $0-\SI{1}{\milli\tesla}$. When addressing these two transitions, the excitation coil needed to be oriented with its main axis along $\mathbf{b}$, in agreement with our model (see main text). The linear polarization of the input light was aligned to be parallel to $\mathbf{D_2}$, so to maximize the absorption.
Our predictions were also confirmed by $T_2^\text{(s)}$ measurements on site I at $\SI{339}{\mega\hertz}$ and $\SI{823}{\mega\hertz}$ respectively. Similarly to site II, values of $T_2^\text{(s)}$ above $\SI{}{\milli\second}$ were found, with a field orientation along $\mathbf{b}$ at $\SI{10}{\milli\tesla}$.
Spectral hole lifetimes above $\SI{300}{\milli\second}$ for temperatures below $\SI{4}{\kelvin}$ assured us that the coherence was not limited by the population lifetime.

The ensemble was first prepared by optical pumping at frequency $\nu_0$ in $\SI{250}{\milli\second}$ with few $\SI{}{\milli\watt}$.
The duration of the $\pi$ pulse in our echo sequence was of $\SI{1.2}{\micro\second}$ and a Rabi frequency up to $\SI{2.5}{\mega\hertz}$ was measured by driving the spin transition at $\SI{655}{\mega\hertz}$ and probing at the same time the transmitted intensity of light on the detector.
The signal generated on the detector was analysed with a Fast Fourier Transform and the area of the peak corresponding to the beat at $|\nu_{LO}-\nu_R|$ was acquired at various time delays $\tau$. The coherence lifetime $T_2^\text{(s)}$ was then extracted by fitting the peak area decays as $I_\text{echo}(\tau)=I_0 \exp(-4\tau/T_2^\text{(s)})$. 

The maxima of coherence lifetimes at non-zero magnetic field were found by optimizing the intensity of the echo at increasingly long delays $\tau$ and correcting accordingly the magnetic field orientation. Similarly, we looked for the highest intensity of the echo when compensating the bias field of the lab while applying small intensities of external field.

The error bars are obtained as follows: the amplitude and width of the echo signal are extracted from a Gaussian fit with 95\% confidence intervals. The corresponding residuals are propagated to find the error on the area of the echo. The latter is in turn propagated on the natural logarithm of the area. For each value of magnetic field, the logarithm of the areas are fitted as a linear function of $\tau$ to find $T_2$. The error on $T_2$ corresponds to a 95\% confidence interval on this last linear fit.

\subsection*{Photon echo}

The optical coherence lifetime $T_2^\text{(o)}$ is measured in an analogous way by preparing a photon echo sequence on a transition between a ground and an excited state. In this case, no preparation of population is needed, since the excited state is empty at the temperature considered, and the echo is readily detectable from the photodiode without RHS. However, we chose to measure an RHS signal at $|\nu_\text{LO}-\nu_0|$ since, in our case, the technique leads to increased sensitivity to the echo amplitude. The duration of the $\pi$ pulse was about $\SI{3}{\micro\second}$. The errors on the coherence lifetimes are obtained as explained in the spin echo section.

\subsection*{Generalization of zero--field ZEFOZ for higher nuclear spins}

We find that the ZEFOZ behavior is generic for all transitions at zero magnetic field under the following sufficient conditions: 
\begin{enumerate}
\item The hyperfine coupling is fully anisotropic (i.e., all three eigenvalues of the coupling tensor A are different and nonzero) to avoid degeneracies in the Hamiltonian; 
\item The electronic spin is $\frac{1}{2}$ and the nuclear spin is half integer.
\end{enumerate}
 This situation leads to a symmetry in the hyperfine coupling and maximally entangled eigenstates. The latter implies a vanishing polarization of the electronic spin and hence a vanishing first order sensitivity to external magnetic fields. Also the polarization of the nuclear spin vanishes due to this symmetry. The ZEFOZ behavior in general disappears when the symmetry is broken, that is, when eigenvalues degenerate, when the nuclear spin is an integer or when the electronic spin is not $\frac{1}{2}$.
 
Let us consider the situation with higher nuclear spin $I = \frac{3}{2}$, coupled with  electronic spin one half. The hyperfine interaction alone leads to eigenstates of the form $\ket{\xi}~= \alpha \ket{\uparrow, \frac{3}{2}} + \beta \ket{\uparrow, - \frac{1}{2}} - \beta \ket{\downarrow, \frac{1}{2}} - \alpha \ket{\downarrow, -\frac{3}{2}}$, where $\alpha$, $\beta$ are coefficients defined by the eigenvalues of the $\mathbf{A}$ tensor. The average electronic polarization $\expval{\mathbf{S}}$ is zero since the wavefunction can be rewritten as $\ket{\xi}~=  \ket{\uparrow}\otimes( \alpha \ket{\frac{3}{2}} + \beta \ket{-\frac{1}{2}}) + \ket{\downarrow}\otimes( - \beta \ket{\frac{1}{2}} - \alpha \ket{-\frac{3}{2}})$, where the weights for $\ket{\uparrow}$ and $\ket{\downarrow}$ components are the same and $\abs{\alpha}^2 + \abs{\beta}^2 = 1/2$. The nuclear polarization $\expval{\mathbf{I}}$ appears to be zero as well, since the projections with the opposite signs appear to have same probabilities giving $\expval{I} = \vqty{\alpha}^2\frac{3}{2} - \vqty{\beta}^2\frac{1}{2} + \vqty{\beta}^2\frac{1}{2} - \vqty{\alpha}^2\frac{3}{2} = 0$. Vanishing $\expval{S}$ and $\expval{I}$ are also found for higher half-integer nuclear spins ($\frac{3}{2}$, $\frac{5}{2}$, $\frac{7}{2}$ etc.). 

For the full description one has to consider the quadrupolar tensor which also contributes to the zero-field splittings. However, the quadrupole interaction $\mathbf{I}\cdot \mathbf{Q}\cdot \mathbf{I}$ does not disturb the symmetry of the wavefunction  for  half-integer $I$. Hence, the ZEFOZ condition persists even in the presence of  the quadrupole interaction. In contrast, if $\mathbf{A}$ and $\mathbf{Q}$ are not aligned and 
$I$ is an integer, the quadrupolar $\mathbf{I}\cdot \mathbf{Q}\cdot \mathbf{I}$  term seems to make the eigenstates even more  susceptible to fluctuations in the $x-y$ plane (in the eigenbasis of $\mathbf{A}$).

\subsection*{Low magnetic field assisted coherence enhancement}

In presence of a relatively small magnetic field (few millitesla) it is still possible to find an orientation which preserves the coherence. In \cite{SpectroYb}, it was found that the $\mathbf{g}$ and $\mathbf{A}$ tensors are highly anisotropic ($g_x \neq g_y \neq g_z$ together with $A_x \neq A_y \neq A_z$) with their strongest component along very close directions. The following calculations can thus be simplified by assuming that the two tensors diagonalize in the same basis, while keeping the $z$ direction as the one along which their components are strongest. Using Eq.~\eqref{eq:Heff}, we can see how the transition sensitivities change with an external static magnetic field $\mathbf{B}$ with ($B_x$, $B_y$, $B_z$) components  by computing the magnitude of the gradient $\mathbf{S}_1$ for one of the transitions:
\begin{equation}
\vqty{\mathbf{S}_1} = 2 \mu_\text{B}^2 \sqrt{\frac{B_z^2 g_z^4}{(A_x+A_y)^2}+\frac{B_y^2g_y^4 A_x^2}{(A_x^2-A_z^2)^2}+\frac{B_x^2 g_x^4 A_y^2}{(A_y^2-A_z^2)^2}}.
\end{equation}
Considering $A_z\gg A_x, A_y$ and $g_z \gg g_x, g_y$, the presence of the product $B_z \, g_z$ on the numerator of the first term and $A_z$ in the denominator of the other terms tells us that to minimize the transition gradient it is convenient to choose a magnetic field orthogonal to the $z$ axis. The expression above can thus be simplified:
\begin{equation}
\vqty{\mathbf{S}_1} = 2 \mu_\text{B}^2 \frac{B}{A_z^2}\sqrt{g_y^4 A_x^2 \sin^2\varphi + g_x^4 A_y^2 \cos^2 \varphi},
\end{equation}
where $\varphi$ is the angle on the plane orthogonal to $z$. Minimizing this expression leads us to a non-zero minimum of the gradient, similarly to a ZEFOZ, with a consequent suppression of decoherence.

\subsection*{Data availability}
 
The data sets generated and/or analysed during the current study are available from the corresponding authors upon reasonable request.

\section*{ACKNOWLEDGEMENTS}

We acknowledge funding from the Swiss FNS NCCR programme Quantum Science Technology (QSIT) and FNS Research Project No 172590, EUs H2020 programme under the Marie Sk\l{}odowska-Curie project QCALL (GA 675662), EU’s FP7 programme under the ERC AdG project MEC (GA 339198), ANR under grant agreement no. 145-CE26-0037-01 (DISCRYS) and Nano’K project RECTUS and the IMTO Cancer AVIESAN (Cancer Plan, C16027HS, MALT).

\subsection*{Competing interests}

The authors declare no competing interests.

\newpage
\clearpage



\section*{Supplementary material}

# Supplementary materials: Simultaneous coherence enhancement of optical and microwave transitions in solid-state electronic spins

Antonio Ortu,[1, *] Alexey Tiranov,[1, *] Sacha Welinski,[2] Florian Fröwis,[1]
Nicolas Gisin,[1] Alban Ferrier,[2, 3] Philippe Goldner,[2] and Mikael Afzelius[1, †]

[1]Groupe de Physique Appliquée, Université de Genève, CH-1211 Genève, Switzerland
[2]Chimie ParisTech, PSL University, CNRS, Institut de Chimie de Paris, 75005 Paris, France
[3]Sorbonne Université, Faculté des Sciences et Ingénierie, UFR 933, Paris, France
(Dated: July 27, 2018)

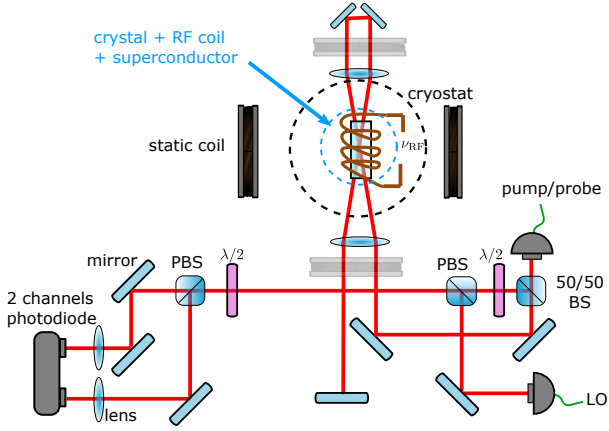

FIG. 1. **Experiment setup**

## I. SETUP

Our setup (excluding beams preparation) is shown in FIG. 1. The pump/probe beams traverse the crystal in a quadruple pass path, to increase the absorption and thus the final signal to noise ratio. The detection part is composed of a balanced heterodyne arrangement, so that we can detect interference with a local oscillator with zero offset. The combination of PBSs and waveplates allows for the alignment of light polarization according to the orientation of the extinction axes of the crystal.

## II. RAMAN HETERODYNE SCATTERING FOR OPTICAL DETECTION OF SPIN ECHO

The first step consists of the creation of a population difference between the two ground states through optical pumping via the $|\phi^-\rangle_e$ state at frequency $\nu_0$. After a time longer than the excited state lifetime, a $\pi/2$ RF pulse at frequency $\nu_{RF}$ is sent to the excitation coil so to create a collective coherent superposition of two hyperfine states in an allowed transition, for example $|\psi^-\rangle_g$ and $|\psi^+\rangle_g$ of optical site II. The inhomogeneous broadening typical of RE doped crystals induces a dephasing of the atomic ensemble, a process that we reverse by a $\pi$ pulse after a time $\tau$ from the $\pi/2$ pulse. Following the rephasing, a collective coherent emission (echo) will take place on the hyperfine transition after another time interval $\tau$. We read out optically the echo by sending at the same time a probe pulse again at $\nu_0$, stimulating a Raman emission on the $|\phi^-\rangle_e \rightarrow |\psi^+\rangle_g$ transition at frequency $\nu_R$. The Raman light is then combined on a beam splitter with a reference beam (local oscillator, LO) from the same laser, but with a frequency $\nu_{LO}$ shifted with respect to the pump/probe beam by a chosen value through a phase modulator. The interference between Raman and LO is detected as a beat at the frequency $|\nu_{LO} - \nu_R|$ in the Fourier spectrum of the signal obtained from a silicon photodiode. A particular strength of this technique is that microwave photons can be detected optically at any frequency without being limited by the bandwidth of the detector, provided that the phase modulator is able to produce at least one sideband on the LO of frequency close enough to the splitting of the hyperfine states $\nu_{RF}$.

The area of the beat signal so acquired, fitted as a function of the delay $\tau$ of the echo sequence, provides the coherence lifetime $T_2$ according to [cite]

$$I_{echo}(\tau) = I_0 e^{-4\tau/T_2}. \tag{1}$$

Some examples of measurements for $T_2$ from spin echo decays are represented in FIG. 2 and FIG. 3.

## III. RHS FOR OPTICAL DETECTION OF SPIN RESONANCES

After a pumping pulse as in the previous section, if we address at the same time the optical and a hyperfine transition, an RHS signal can be detected showing the presence of a microwave resonance. By scanning the frequency $\nu_{RF}$, we can study the variation of the transition frequencies as we change the external magnetic field. FIG. 4a shows an example of scan for the 655 MHz transition of site II and small magnetic field values along $\mathbf{D_2}$. We can distinguish the two magnetic sub-sites as they split in frequency when few millitesla are applied, while they are equivalent at zero field. FIG. 4b represents a similar scan around an angle of decoherence suppression on the $\mathbf{D_2} - \mathbf{b}$ plane. The transition frequency has a minimum at this angle, consistently with our prediction that the components of the gradient orthogonal to the applied magnetic field should vanish. The two magnetic

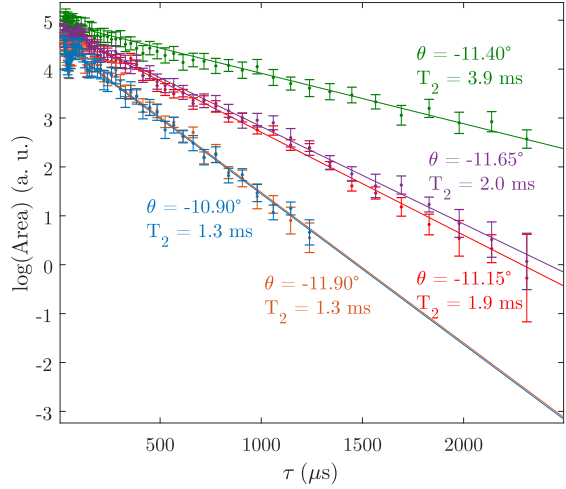

FIG. 2. **Spin echo decay, site II.** Examples of spin echo area measurements for varying delay $\tau$, at different values of $\theta$ around a direction of magnetic field of maximum decoherence suppression for $\varphi = -70°$ and $|B| = 5$ mT.

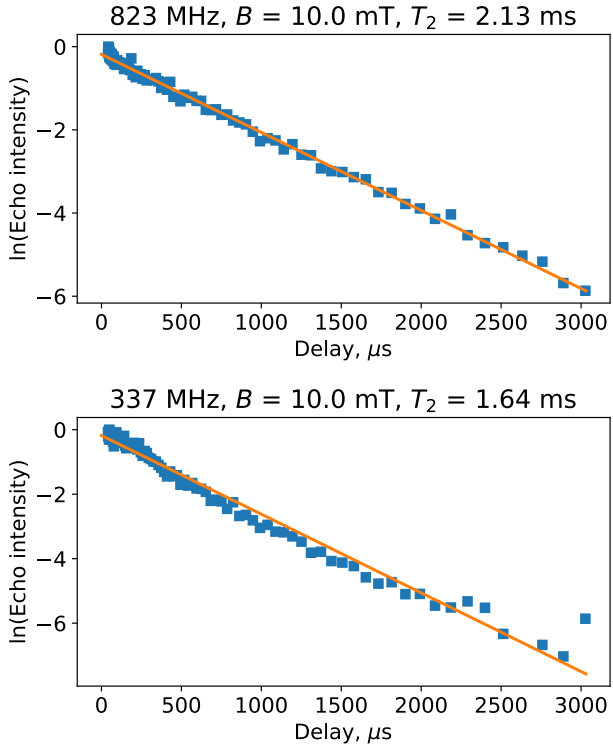

FIG. 3. **Spin echo decay, site I.** Example of spin echo decay measurement with related fit for $T_2$, showing compatible result with data from site II.

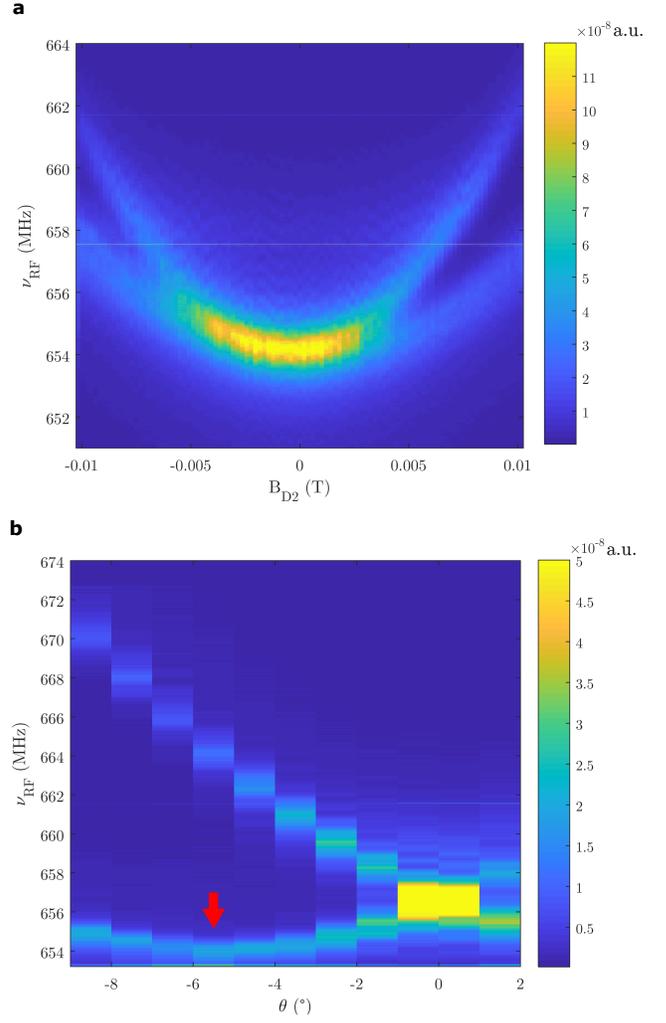

FIG. 4. **RHS spin resonance detection. a** RHS signal for a scan of the microwave excitation frequency $\nu_0$RF and magnetic field along $D_2$. **b** RHS signal around the angle of decoherence suppression, the corresponding transition is indicated by the red arrow. The magnetic field amplitude of 7 mT applied on the $\mathbf{D_2} - \mathbf{b}$ plane was used.

sub-sites appear well separated in frequency for most of the scanned range, but become again equivalent when the applied field is orthogonal to $\mathbf{b}$ ($\theta = 0°$).



\end{document}